\documentclass[11pt]{article}

\usepackage{lineno,hyperref}
\modulolinenumbers[5]

\usepackage{amsmath}
\usepackage{amssymb}
\usepackage{amsthm}
\usepackage{tikz}
\usepackage{pgfplots}
\usepackage{graphicx}
\usepackage{slashbox}
\usepackage{caption}
\usepackage{subcaption}
\usepackage{xcolor}
\usepackage{balance}
\allowdisplaybreaks  
\usepackage{cite}
\usepackage{algorithm}
\usepackage[noend]{algcompatible}
\usepackage{mathtools}
\usepackage{subcaption}
\usepackage{hyperref}
\usepackage{comment}
\usepackage{tabularx}
\usepackage{bbm}
\usepackage{thmtools}

\allowdisplaybreaks

\renewcommand{\thmcontinues}[1]{cont.}

\usepackage{array}
\usepackage{authblk}
\usepackage{setspace}
\usepackage{fullpage,etoolbox}
\newtheorem{theorem}{Theorem}
\newtheorem{lemma}{Lemma}
\newtheorem{assumption}{Assumption}

\newtheorem{remark}{Remark}
\newtheorem{problem}{Problem} % definition numbers are dependent on theoremnumbers

\newcommand{\vertiii}[1]{{\vert\kern-0.25ex\vert\kern-0.25ex\vert #1\vert\kern-0.25ex\vert\kern-0.25ex\vert}}

\newtoggle{draft}
\togglefalse{draft}

\title{Conformal Prediction for STL Runtime Verification}
\author[1]{Lars Lindemann\thanks{Lars Lindemann and Xin Qin contributed equally.}}
\author[1]{Xin Qin$^*$}
\author[1]{Jyotirmoy V. Deshmukh}
\author[2]{George J. Pappas}
\affil[1]{Department of Computer Science, University of Southern California}
\affil[2]{Department of Electrical and Systems Engineering, University of Pennsylvania}

\begin{document}

\maketitle

%%%%%%%%%%%%%%%%%%%%%%%%%%%%%%%%%%%%%%%%%%%%%%%%%%%%%%%%%%%%%%%%%%%%%%%%%%%%%%%%
\begin{abstract}
We are interested in predicting failures of cyber-physical systems during their operation. Particularly, we consider stochastic systems and signal temporal logic specifications, and we want to calculate the probability that the current system trajectory violates the specification. The paper presents two predictive runtime verification algorithms that predict future system states from the current observed system trajectory. As these predictions may not be accurate, we construct prediction regions that quantify prediction uncertainty by using conformal prediction,  a statistical tool for uncertainty quantification. Our first algorithm directly constructs a prediction region for the satisfaction measure of the specification so that we can predict specification violations with a desired confidence. The second algorithm constructs prediction regions for future system states first, and uses these  to obtain a prediction region for the satisfaction measure. To the best of our knowledge, these are the first formal guarantees for a predictive runtime verification algorithm that applies to widely used trajectory predictors such as RNNs and LSTMs, while being computationally simple and making no assumptions on the underlying distribution. We present  numerical experiments of an F-16 aircraft and a self-driving car.  
\end{abstract}

%%%%%%%%%%%%%%%%%%%%%%%%%%%%%%%%%%%%%%%%%%%%%%%%%%%%%%%%%%%%%%%%%%%%%%%%%%%%%%%%

\section{Introduction}
\label{sec:introduction}

Cyber-physical systems may be subject to a small yet non-zero failure probability, especially when using data-enabled perception and decision making capabilities, e.g., self-driving cars using high-dimensional sensors. Rare yet catastrophic system failures hence have to be anticipated. In this paper, we aim to detect system failures with high confidence early on during the operation of the system.

Verification aims to check the correctness of a system against specifications expressed in mathematical logics, e.g., linear temporal logic \cite{pnueli1977temporal} or signal temporal logic (STL) \cite{maler2004monitoring}. Automated verification tools were developed for deterministic systems, e.g.,  model checking \cite{baier,clarke1997model} or theorem proving \cite{shoukry2017smc,sheeran2000checking}. Non-deterministic system verification was studied using probabilistic model checking \cite{bianco1995model,hansson1994logic,kwiatkowska2011prism,jackson2021formal} or statistical model checking \cite{younes2002probabilistic,younes2006statistical,legay2019,legay2010statistical}.  Such offline verification techniques have been  applied to verify cyber-physical systems, e.g., autonomous race cars \cite{ivanov2019verisig,ivanov2020case,lindemann2022risk}, cruise controller and emergency braking systems \cite{tran2020nnv,tran2019safety}, autonomous robots \cite{sun2019formal}, or aircraft collision avoidance systems \cite{bak2021second,bak2022neural}. 

These verification techniques, however, are: 1) applied to a system model that may not  capture the system sufficiently well, and 2) performed offline and not during the runtime of the system. We may hence certify a system to be safe a priori (e.g., with a probability of $0.99$), but during the system's runtime we may observe an unsafe system realization (e.g., belonging to the fraction of $0.01$ unsafe realizations). Runtime verification aims to detect unsafe system realizations by using online monitors to observe the current  realization (referred to as prefix) to determine if all  extensions of this partial realization (referred to as suffix) either satisfy or violate the specification, see \cite{bauer2011runtime,leucker2009brief,cassar2017survey} for deterministic and \cite{wilcox2010runtime,sistla2011runtime,jaeger2020statistical} for non-deterministic systems. The verification answer can be inconclusive when not all suffixes are satisfying or violating. Predictive runtime verification instead predicts suffixes from the prefix to obtain a verification result more reliably and quickly \cite{babaee2018mathcal,yoon2019predictive,qin2020clairvoyant}. 

\begin{figure*}
\centering
\includegraphics[scale=0.24]{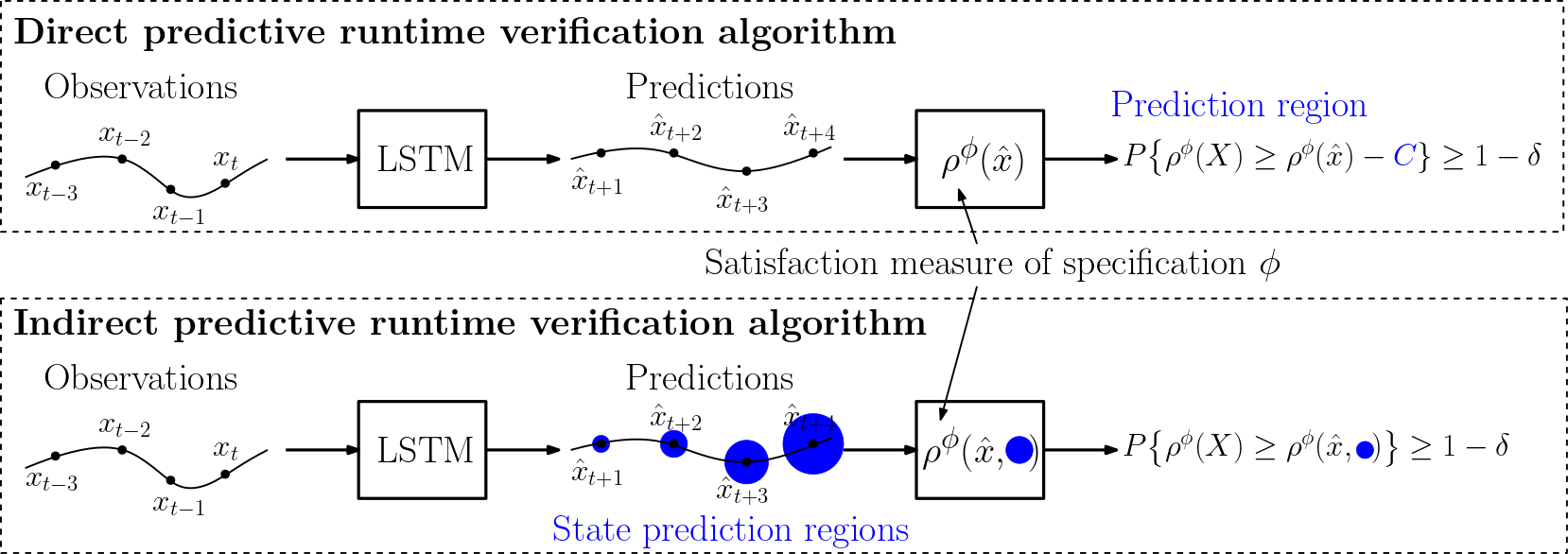}
\caption{Overview of the proposed STL predictive runtime verification algorithms. Both algorithms use past observations $(x_0,\hdots,x_t)$ to obtain state predictions $(\hat{x}_{t+1},\hat{x}_{t+2},\hdots)$. The direct algorithm calculates the satisfaction measure $\rho^\phi(\hat{x})$ of the specification $\phi$ based on these predictions, and obtains a prediction region $C$ for the unknown satisfaction measure $\rho^\phi(x)$ using conformal prediction. The indirect method obtains prediction regions for the unknown states $x_{t+1},x_{t+2},\hdots$ using conformal prediction first, and then obtains a lower of the unknown satisfaction measure $\rho^\phi(x)$ based on the state prediction regions. }
\label{fig:1}
\end{figure*}

We are interested in the predictive runtime verification of a stochastic system, modeled by an unknown distribution $\mathcal{D}$, against a system specification $\phi$ expressed in STL. Particularly, we want to calculate the probability that the current system execution violates the specification based on the current observed trajectory, see Figure \ref{fig:1}. To the best of our knowledge, existing predictive runtime verification algorithms do not provide formal correctness guarantees unless restrictive assumptions are placed on the prediction algorithm or the underlying distribution $\mathcal{D}$. We allow the use of complex prediction algorithms such as recurrent neural networks (RNNs) and long short-term memory (LSTM) networks, while making no assumptions on $\mathcal{D}$.  Our contributions are as follows:
\begin{itemize}
\item We present two  predictive runtime verification algorithms that are illustrated in Figure \ref{fig:1} and that use: i) trajectory predictors to predict future system states, and ii) conformal prediction to quantify  prediction uncertainty. 
\item  We show that our algorithms enjoy valid verification guarantees, i.e., the verification answer is correct with a user-defined confidence, with minimal assumptions on the predictor and the underlying distribution $\mathcal{D}$. 
%\item \textcolor{red}{We show how to use our results to assess the quality of the LSTM online. This information can be important for online decision making.}
\item We provide realistic empirical validation of our approach of an F-16 aircraft and a self-driving car, and compare the two proposed runtime verification algorithms.
\end{itemize}

\subsection{Related Work}

\textbf{Statistical model checking. } 
Statistical model checking is a lightweight alternative to computationally expensive probabilistic model checking used to verify black-box systems \cite{younes2002probabilistic,younes2006statistical,legay2019,legay2010statistical}. The idea is to sample system trajectories and use statistical tools to get valid verification guarantees. Statistical model checking has gained popularity  due to the complexity of modern machine learning architectures for which it is difficult to obtain meaningful, i.e., not overly conservative, analytical results. 

We focus on signal temporal logic (STL) as a rich specification language \cite{maler2004monitoring} that admits robust semantics  to quantify how robustly a system satisfies a specification spatially and/or temporally \cite{fainekos2009robustness,donze2,alena_journal}. Statistical model checking under STL specifications was first considered in \cite{bartocci2013robustness,bartocci2015system}, while \cite{salamati2020data,salamati2021data,jackson2021formal} proposed a combination of a statistical and a model-based approach. The authors in \cite{Wang2019,zarei2020statistical,Wang2021,Roohi2017} use statistical testing to derive high confidence bounds on the probability of a cyber-physical system satisfying an STL specification.  In \cite{Chapman2021,lindemann2022risk,lindemann2022temporal,akella2022scenario,akella2022sample} risk verification algorithms were proposed using mathematical notions of risk.

\textbf{Predictive Runtime Verification. } Runtime verification complements system verification by observing the current system execution (prefix) to determine if all extensions (suffixes) either satisfy or violate  the specification  \cite{bauer2011runtime,leucker2009brief,cassar2017survey,wilcox2010runtime,sistla2011runtime,jaeger2020statistical}. Runtime verification is an active research area \cite{lukina2021into,ruchkin2022confidence,boursinos2021assurance}, and  algorithms were recently proposed  for verifying STL properties and hyperproperties in \cite{gressenbuch2021predictive,deshmukh2017robust,selvaratnam2022mitl} and  \cite{finkbeiner2019monitoring,hahn2019algorithms}, respectively. While the verification result in runtime verification can be inconclusive, predictive runtime verification predicts a set of possible suffixes (e.g., a set of potential trajectories) to provide a verification result more reliably and quickly. In \cite{yoon2019predictive,pinisetty2017predictive,yu2022online,yu2022model,althoff2014online,koschi2018set}, knowledge of the system  is assumed to obtain predictions of system trajectories. However, the system is not always exactly known so that in \cite{ferrando2021incrementally,babaee2019accelerated,babaee2018mathcal} a system model is learned first, while in \cite{yoon2021predictive,chou2020predictive,qin2020clairvoyant,tan2022trajectory} future system trajectories are predicted from past observed data using trajectory predictors. To the best of our knowledge, none of these works provide valid verification guarantees unless the system is exactly known or strong assumptions are placed on the prediction algorithm.

\textbf{Conformal Prediction. }   Conformal prediction was introduced in \cite{vovk2005algorithmic,shafer2008tutorial} as a statistical tool to quantify uncertainty of prediction algorithms. In \cite{luo2021sample}, conformal prediction was used to obtain guarantees on the false negative rate of an online monitor. Conformal prediction was  used for verification of STL properties in \cite{qin2022statistical} by learning a predictive model of the STL semantics. For reachable set prediction, the authors in  \cite{bortolussi2019neural,bortolussi2021neural,cairoli2021neural} used conformal prediction to quantify uncertainty of a predictive runtime monitor that predicts reachability of safe/unsafe states. However, the works in \cite{qin2022statistical,bortolussi2019neural,bortolussi2021neural,cairoli2021neural} train task-specific predictors while we use task-independent trajectory predictors to predict future system states from which we infer information about the satisfaction of the task. This is significant as no expensive retraining is required when the specification changes. The authors of the work in \cite{cairoli2022conformal}, which appeared concurrently with our paper, also consider predictive runtime verification under STL specifications. Similar to our work, they provide probabilistic guarantees for the quantitative semantics of STL, but consider a different runtime verification setting in which systems have to be Markovian. Again, their predictors are  task-specific while our predictors are task-independent so that we avoid expensive retraining when specifications change.

\section{Problem Formulation}
\label{sec:rnn}

Let $\mathcal{D}$ be an unknown distribution over system trajectories that describe our system, i.e., let $X:=(X_0,X_1\hdots)\sim \mathcal{D}$ be a random trajectory where $X_\tau$ denotes the state of the system at time $\tau$ that is drawn from $\mathbb{R}^n$.  Modeling stochastic  systems by a distribution $\mathcal{D}$ provides great flexibility, and $\mathcal{D}$ can generally describe the motion of Markov decision processes. It can capture stochastic systems whose trajectories follow the recursive update equation $X_{\tau+1}=f(X_\tau,w_\tau)$ where $w_\tau$ is a random variable and where the (unknown) function $f$ describes the system dynamics. Stochastic systems can describe the behavior of engineered systems such as robots and autonomous systems, e.g., drones or self-driving cars, but they can also describe weather patterns, demographics, and human motion. We use lowercase letters $x_\tau$ for realizations of the random variable $X_\tau$. We make no assumptions on the distribution  $\mathcal{D}$, but assume availability of training and calibration data drawn from $\mathcal{D}$. 
\begin{assumption}\label{ass1}
 	We have access to $K$ independent realizations $x^{(i)}:=(x_0^{(i)},x_1^{(i)},\hdots)$ of the distribution $\mathcal{D}$ that are collected in the dataset $D:=\{x^{(1)},\hdots,x^{(K)}\}$.
 \end{assumption}

\textbf{Informal Problem Formulation.} Assume now that we are  given a specification $\phi$ for the stochastic system $\mathcal{D}$, e.g., a safety or performance specification defined over the states $X_\tau$ of the system. In  ``offline'' system verification, e.g., in statistical model checking, we are interested in calculating the probability that $(X_0,X_1,\hdots)\sim \mathcal{D}$ satisfies the specification. In runtime verification, on the other hand, we have already observed the partial realization $(x_0,\hdots,x_t)$ of $(X_0,\hdots,X_t)$ online at time $t$, and we want to use this information to calculate the probability that $(X_0,X_1,\hdots)\sim \mathcal{D}$  satisfies the specification.\footnote{We note that we consider  unconditional probabilities in this paper.}  In this paper, we use predictions $\hat{x}_{\tau|t}$ of future states  $X_\tau$ for this task in a predictive runtime verification approach.  

\begin{figure}
\centering
\includegraphics[scale=0.2]{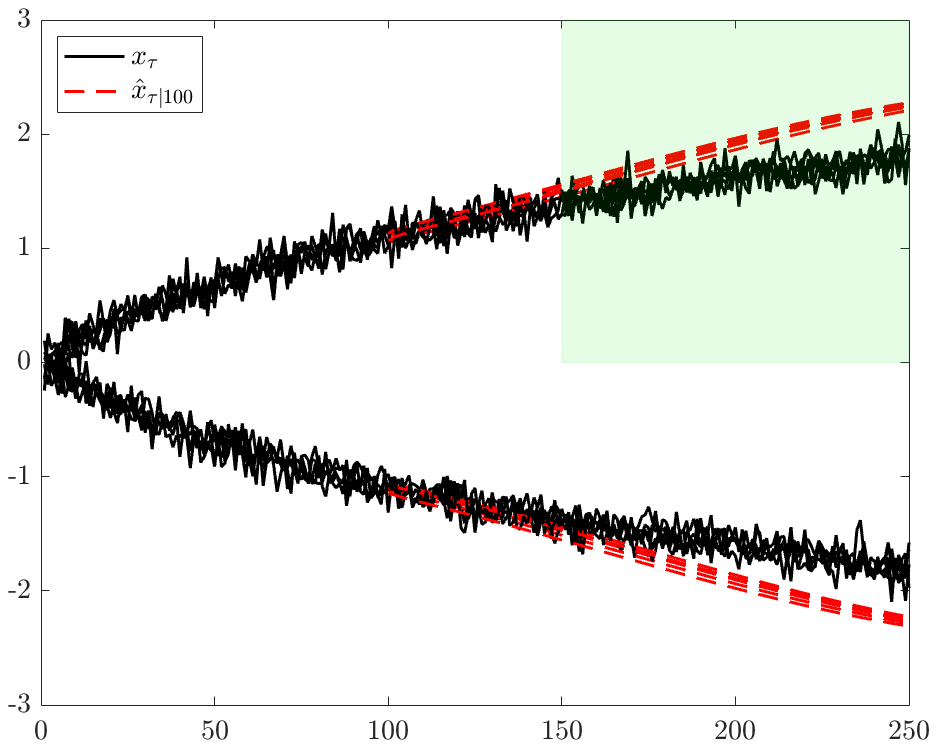}\hspace{0.5cm}
\includegraphics[scale=0.2]{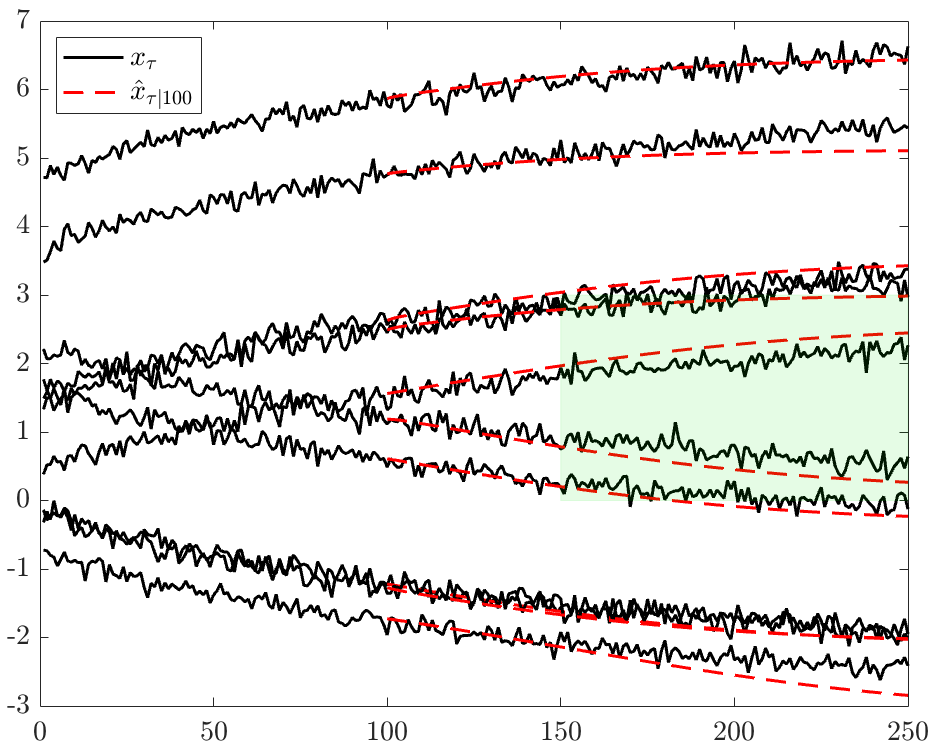}
\caption{Ten realizations of two stochastic systems (solid lines) and corresponding LSTM predictions at time $t:=100$ (red dashed lines). The specification is that trajectories should be within the green box between $150$ and $250$ time units.}
\label{fig:11}
\end{figure}

While in ``offline'' verification all realizations of $\mathcal{D}$ are taken into account, only a subset of these are relevant in  runtime verification. One hence gets different types of verification guarantees, e.g., consider a stochastic system $(X_0,X_1,\hdots)\sim \mathcal{D}$ of which we have plotted ten realizations in Figure~\ref{fig:11} (left). In an offline approach, this system satisfies the specification $\inf_{\tau\in[150,250]} X_\tau\in[0,3]\ge 0$  with a probability of $0.5$. However, given an observed partial realization $(x_1,\hdots,x_{100})$, we are  able to give a better answer. In this case, we used LSTM predictions $\hat{x}_{\tau|100}$ (red dashed lines), to more confidently say if the specification is satisfied. While the stochastic system in Figure \ref{fig:11} (left) has a simple structure, the same task for the stochastic system in Figure \ref{fig:11} (right) is already more challenging.

 %We emphasize that specifications are here  defined over the system and the RNN, i.e., over system states $x_\tau$ and RNN predictions $\hat{x}_{\tau|t}$. We next provide some motivating examples.
%\begin{itemize}
%    \item \textbf{System specifications.} We may want that the system satisfies performance and safety specifications, i.e., visiting a set of locations and avoiding dangerous areas. Given the observed partial sequence $(x_0,\hdots,x_t)$, what is the probability that $(x_0,\hdots,x_t,X_{t+1},\hdots,X_{t+H})$ satisfies such a system specification?
%    \item \textbf{Accuracy in forecasting.} We may want that the $k$-step ahead prediction error is smaller than a  threshold of $\epsilon>0$, i.e., that $\|X_{t+k}-\hat{x}_{t+k|t}\|\le \epsilon$ holds. In robotics where $x$ can denote the position of a mobile robot, we may additionally want that the $k$-step ahead prediction error is always, i.e., for all $k\in\{1,\hdots,H\}$, smaller than $\epsilon$ for collision avoidance objectives. 
%    \item \textbf{Side information in forecasting.} We want that predictions respects side information, e.g., from physical prior knowledge. For instance, when a car is predicted to travel at $30$ km/h, the predictions can not change to $100$ km/h abruptly at the next time step.
%\end{itemize}

\subsection{Signal Temporal Logic}
\label{sec:STL}
To express  system specifications, we use signal temporal logic (STL). Let $x:=(x_0,x_1,\hdots)$ be a discrete-time signal, e.g., a realization of the stochastic system $(X_0,X_1,\hdots)$. The atomic elements of STL are predicates that are functions $\mu:\mathbb{R}^n\to\{\text{True},\text{False}\}$. For convenience, the predicate $\mu$ is often defined via a predicate function $h:\mathbb{R}^n\to\mathbb{R}$ as $\mu(x_\tau):=\text{True}$ if $h(x_\tau)\ge 0$ and $\mu(x_\tau):=\text{False}$ otherwise. The syntax of STL is recursively defined as 
\begin{align}\label{eq:full_STL}
\phi \; ::= \; \text{True} \; | \; \mu \; | \;  \neg \phi' \; | \; \phi' \wedge \phi'' \; | \; \phi'  U_I \phi'' \; | \; \phi' \underline{U}_I \phi'' \,
\end{align}
where $\phi'$ and $\phi''$ are STL formulas. The Boolean operators $\neg$ and $\wedge$ encode negations (``not'') and conjunctions (``and''), respectively. The until operator $\phi' {U}_I \phi''$ encodes that $\phi'$ has to be true from now on until $\phi''$ becomes true at some future time within the time interval $I\subseteq \mathbb{R}_{\ge 0}$. The since operator encodes that $\phi''$ was true at some past time within the time interval $I$ and since then $\phi'$ is true. We can further derive the operators for disjunction ($\phi' \vee \phi'':=\neg(\neg\phi' \wedge \neg\phi'')$), eventually ($F_I\phi:=\top U_I\phi$), once ($\underline{F}_I\phi:=\top \underline{U}_I\phi$), always ($G_I\phi:=\neg F_I\neg \phi$), and historically ($\underline{G}_I\phi:=\neg \underline{F}_I\neg \phi$).

To determine if a signal $x$ satisfies an STL formula $\phi$ that is enabled at time $\tau_0$, we can define the semantics as a relation $\models$, i.e.,  $(x,\tau_0) \models\phi$ means that $\phi$ is satisfied. While the STL semantics are fairly standard \cite{maler2004monitoring}, we recall them in Appendix~\ref{app:STL}. Additionally, we can define robust (sometimes referred to as quantitative) semantics $\rho^{\phi}(x,\tau_0)\in\mathbb{R}$ that indicate how robustly the formula $\phi$ is satisfied or violated \cite{donze2,fainekos2009robustness}, see Appendix~\ref{app:STL}. Larger and positive values of $\rho^{\phi}(x,\tau_0)$ hence indicate that the specification is satisfied more robustly. Importantly, it holds that $(x,\tau_0)\models \phi$ if $\rho^\phi(x,\tau_0)>0$. We make the following assumption on the class of STL formulas in this paper.
\begin{assumption}\label{ass2}
  We consider bounded STL formulas $\phi$, i.e., all time intervals $I$ within the formula $\phi$ are bounded.
 \end{assumption}
 
 Satisfaction of bounded STL formulas can be decided by finite length signals \cite{sadraddini2015robust}. The minimum length is indicated by the formula length $L^\phi$, i.e., knowledge of $(x_0,\hdots,x_{\tau_0+L^\phi})$ is enough to determine if $(x,\tau_0) \models\phi$. We recall the definition of $L^\phi$ in Appendix \ref{app:STL}.

\subsection{Trajectory Predictors}
\label{sec:tra_pred}
Given an observed partial sequence $(x_0,\hdots,x_t)$ at the current time $t\ge 0$,
we want to predict the states $(x_{t+1},\hdots,x_{t+H})$ for a prediction horizon of $H>0$. Our runtime verification algorithm is in general compatible with any trajectory prediction algorithm.  Assume therefore that \textsc{Predict} is a measurable function that maps observations $(x_0,\hdots,x_t)$ to predictions $(\hat{x}_{t+1|t},\hdots,\hat{x}_{t+H|t})$ of $(x_{t+1},\hdots,x_{t+H})$. 

Trajectory predictors are typically learned. We therefore split the dataset $D$ into training and calibration datasets $D_\text{train}$ and $D_\text{cal}$, respectively, and  learn \textsc{Predict} from $D_{\text{train}}$. 
 
A specific example of \textsc{Predict} are recurrent neural networks (RNNs) that have shown good performance \cite{lipton2015critical,rudenko2020human}.  For $\tau\le t$, the recurrent structure of an RNN is given as 
\begin{align*}
     a_\tau^1&:=\mathcal{A}(x_\tau,a_{\tau-1}^1), \;\;\;\;\;\;\;\;\;\;\;\; \\
     a_\tau^i&:=\mathcal{A}(x_\tau,a_{\tau-1}^{i},a_\tau^{i-1}), \;\;\;\; \forall i\in\{2,\hdots,d\}\\
     y_{\tau+1|\tau}&:=\mathcal{Y}(a_\tau^d),
 \end{align*}
 where $x_\tau$ is the input that is sequentially applied to the RNN and where  $\mathcal{A}$ is a function that can parameterize different types of RNNs, e.g., LSTMs  \cite{hochreiter1997long}. Furthermore, $d$ is the RNN's depth and $a_\tau^1,\hdots,a_\tau^d$ are the hidden states. The output $y_{t+1|t}:=(\hat{x}_{t+1|t},\hdots,\hat{x}_{t+H|t})$ provides an estimate of $(x_{t+1},\hdots,x_{t+H})$ via the function $\mathcal{Y}$ which typically parameterizes a linear last layer.

\subsection{Predictive Runtime Verification}

We recall that $(x_0,x_1,\hdots)$ denotes a realization of $X:=(X_0,X_1,\hdots)\sim\mathcal{D}$.
Assume that we have observed $x_{\text{obs}}:=(x_0,\hdots,x_t)$ at time $t$, i.e., all states up until time $t$ are known, while the realizations of $x_{\text{un}}:=(x_{t+1},x_{t+2},\hdots)$ are not known yet. Consequently, we have that ${X}:=(X_{\text{obs}},X_{\text{un}})$.\footnote{For convenience, we chose the notations of $X_{\text{obs}}$, $X_{\text{un}}$, and  ${X}$ that do not explicitly reflect the dependence on the current time $t$.} In this paper, we are interested in calculating the probability that  $({X},\tau_0) \models \phi$ as formally stated next.\footnote{We remark that the semantics and the robust semantics are measurable so that probabilities over these functions are well defined \cite{lindemann2022risk,bartocci2015system}.}

\begin{problem}\label{prob1}
Given a distribution $(X_0,X_1,\hdots)\sim\mathcal{D}$, the current time $t$, the  observations $x_{\text{obs}}:=(x_0,\hdots,x_t)$, a bounded STL formula $\phi$ that is enabled at $\tau_0$, and a failure probability $\delta\in(0,1)$, determine if $P\big((X,\tau_0)\models \phi\big)\ge 1-\delta$ holds.
\end{problem}

Several comments are in order. Note that we use the system specification $\phi$ (and not its negation $\neg\phi$)  to determine if $\phi$ is satisfied. From $P\big((X,\tau_0)\models \phi\big)\ge 1-\delta$, we can infer that $P\big((X,\tau_0)\models \neg \phi\big)\le \delta$, i.e., we get an upper bound on the probability that the specification is violated. We further remark that, as a byproduct of our solution to Problem \ref{prob1}, we obtain a probabilistic lower bound $\bar{C}\in\mathbb{R}$ on the robust semantics $\rho^\phi(X,\tau_0)$, i.e., so that $P\big(\rho^\phi(X,\tau_0)\ge \bar{C}\big)\ge 1-\delta$.

We would like to point out two special instances of Problem~\ref{prob1}. When $\tau_0:=0$, we recover the ``standard'' runtime verification problem in which a specification is enabled at time zero, such as in the example $\inf_{\tau\in[150,250]} x_\tau\in[0,3]$ shown in Figure \ref{fig:11}. When $\tau_0:=t$, the current time coincides with the time the specification is enabled. This may, for instance, be important when monitoring the current quality of a system, e.g., when monitoring the output of a neural network used for perception in autonomous driving.

%Problem \ref{prob1} is a predictive runtime verification problem when $T<t$, while it is a predictive online verification problem when $T=t$. For instance, the specification  in  was imposed at the time $T=0$, while the current time was $t=100$. Instead, considering $\inf_{\tau\in[150,250]} x_{T+\tau}\in[0,3]$ with $T:=t$ would be  predictive online monitoring. While these are just semantic differences, these two problem classes have different use cases in practice.  

\section{Conformal Prediction for Predictive Runtime Verification}

In this section, we first provide an introduction to conformal prediction for uncertainty quantification. We then propose two predictive runtime verification algorithms to solve Problem \ref{prob1}. We refer to these algorithms as direct and indirect. This naming convention is motivated as the direct method applies conformal prediction directly to obtain a prediction region for the robust semantics $\rho^{\phi}(X,\tau_0)$. The indirect method uses conformal prediction to get prediction regions for future states $X_\tau$ first, which are subsequently used indirectly to obtain a prediction region for $\rho^{\phi}(X,\tau_0)$, see Figure \ref{fig:11}.

\subsection{Introduction to Conformal Prediction}
\label{sec:intro_conf}
Conformal prediction was introduced  in \cite{vovk2005algorithmic,shafer2008tutorial} to obtain valid prediction regions for complex prediction algorithms, i.e., neural networks, without making assumptions on the underlying distribution or the prediction algorithm \cite{angelopoulos2021gentle,zeni2020conformal,lei2018distribution,tibshirani2019conformal,cauchois2020robust}.  

We first provide a brief introduction to conformal prediction. Let $R^{(0)},\hdots,R^{(k)}$ be $k+1$ independent and identically distributed random variables. The variable $R^{(i)}$ is usually referred to as the {nonconformity score}. In supervised learning, it may be defined as $R^{(i)}:=\|Y^{(i)}-\mu(X^{(i)})\|$ where the predictor $\mu$ attempts to predict an output  $Y^{(i)}$ based on an input $X^{(i)}$.  A large nonconformity score indicates a poor predictive model. 

Our goal is to obtain a prediction region for $R^{(0)}$ based on $R^{(1)},\hdots,R^{(k)}$, i.e., the random variable $R^{(0)}$ should be contained within the prediction region with high probability.  Formally, given a failure probability $\delta\in (0,1)$, we want to construct a valid prediction region $C$ that depends on $R^{(1)},\hdots,R^{(k)}$ such that 
\begin{align*}
    P(R^{(0)}\le C)\ge 1-\delta.
\end{align*}

As $C$ depends on $R^{(1)},\hdots,R^{(k)}$, note that the probability measure $P$ is defined over the product measure of $R^{(0)},\hdots,R^{(k)}$. This is an important observation as conformal prediction guarantees marginal coverage but not conditional coverage, see \cite{angelopoulos2021gentle} for a detailed discussion.  By a surprisingly simple quantile argument, see \cite[Lemma 1]{tibshirani2019conformal}, one can obtain $C$ to be the $(1-\delta)$th quantile of the empirical distribution of the values $R^{(1)},\hdots,R^{(k)}$ and $\infty$. By assuming that $R^{(1)},\hdots,R^{(k)}$ are sorted in non-decreasing order, and by adding $R^{(k+1)}:=\infty$, we can equivalently obtain $C:=R^{(p)}$ where $p:=\lceil (k+1)(1-\delta)\rceil$, i.e., $C$ is the $p$th smallest nonconformity score.

\subsection{Direct STL Predictive Runtime Verification}

Recall that we can obtain predictions $\hat{x}_{\tau|t}$ of $x_\tau$ for all future times $\tau>t$ using the \textsc{Predict} function. However, the predictions $\hat{x}_{\tau|t}$ are only point predictions that are not sufficient to solve Problem \ref{prob1} as they do not contain any information about the uncertainty of $\hat{x}_{\tau|t}$. 

We first propose a solution  by a direct application of conformal prediction. Let us therefore define $H:=\tau_0+L^\phi-t$ as the maximum prediction horizon that is needed to estimate the satisfaction of the bounded STL specification $\phi$. Define now the predicted trajectory
  \begin{align}\label{eq:pred_traj}
      \hat{x}:=(x_{\text{obs}},\hat{x}_{t+1|t},\hdots,\hat{x}_{t+H|t})
  \end{align}
  which is the concatenation of the current observations $x_{\text{obs}}$ and the predictions of future states $\hat{x}_{t+1|t},\hdots,\hat{x}_{t+H|t}$. For an a priori fixed failure probability $\delta\in(0,1)$, our goal is  to directly construct a prediction region defined by a constant $C$ so that 
\begin{align}\label{eq:pred_direct}
    P\big(\rho^{\phi}(\hat{x},\tau_0)-\rho^{\phi}(X,\tau_0) \le C  \big)\ge 1-\delta.
\end{align} 
Note that $\rho^{\phi}(\hat{x},\tau_0)$ is the predicted robust semantics for the specification $\phi$ that we can calculate at time $t$ based on the observations $x_{\text{obs}}$ and the predictions $\hat{x}_{t+1|t},\hdots,\hat{x}_{t+H|t}$. Now, if equation \eqref{eq:pred_direct} holds, then we know that $\rho^{\phi}(\hat{x},\tau_0)>C$ is a sufficient condition for $P(\rho^{\phi}(X,\tau_0)>0)\ge 1-\delta$ to hold.

To obtain the constant $C$, we thus consider the nonconformity score $R:=\rho^{\phi}(\hat{x},\tau_0)-\rho^{\phi}(X,\tau_0)$.  In fact, let us compute the nonconformity score for each calibration trajectory $x^{(i)}\in D_\text{cal}$ as 
\begin{align*}
    R^{(i)}:=\rho^{\phi}(\hat{x}^{(i)},\tau_0)-\rho^{\phi}(x^{(i)},\tau_0)
\end{align*}
where $\hat{x}^{(i)}:=(x_{\text{obs}}^{(i)},\hat{x}_{t+1|t}^{(i)},\hdots,\hat{x}_{t+H|t}^{(i)})$ resembles equation \eqref{eq:pred_traj}, but now defined for the calibration trajectory $x^{(i)}$.\footnote{This means that $\hat{x}^{(i)}$  is the concatenation of the observed calibration trajectory  $x_{\text{obs}}^{(i)}:=(x_0^{(i)},\hdots,x_t^{(i)})$ and the predictions $\hat{x}_{t+1|t}^{(i)},\hdots,\hat{x}_{t+H|t}^{(i)}$  obtained from  $x_{\text{obs}}^{(i)}$.} A positive nonconformity score $R^{(i)}$ indicates that our predictions are too optimistic, i.e., the predicted robust semantics $\rho^{\phi}(\hat{x}^{(i)},\tau_0)$ is greater than the actual robust semantics $\rho^{\phi}(x^{(i)},\tau_0)$ obtained when using the ground truth calibration trajectory $x^{(i)}$. Conversely, a negative value of $R^{(i)}$ means that our prediction are too conservative.

We can now directly obtain a constant  $C$ that makes equation \eqref{eq:pred_direct} valid, and use this $C$  to solve Problem \ref{prob1}, by a direct application of \cite[Lemma 1]{tibshirani2019conformal}. Therefore assume, without loss of generality, that the values of $R^{(i)}$ are sorted in non-decreasing order and let us add  $R^{(|D_\text{cal}|+1)}:=\infty$ as the $(|D_\text{cal}|+1)$th value. 
  
\begin{theorem}\label{thm:1}
Given a distribution $(X_0,X_1,\hdots)\sim\mathcal{D}$, the current time $t$, the observations $x_{\text{obs}}:=(x_0,\hdots,x_t)$, a bounded STL formula $\phi$ that is enabled at $\tau_0$, the dataset $D_\text{cal}$, and a failure probability $\delta\in(0,1)$. Then the prediction region in equation \eqref{eq:pred_direct} is valid with $C$
defined as
\begin{align}\label{eq:thm_1}
    C:=R^{(p)}\;\;\text{where}\;\;  p:=\big\lceil (|D_\text{cal}|+1)(1-\delta)\big\rceil,
\end{align}
and it holds that $P\big((X,\tau_0)\models \phi\big)\ge 1-\delta$ if $\rho^{\phi}(\hat{x},\tau_0)> C$.

\begin{proof}
The nonconformity scores $R^{(i)}$ are independent and identically distributed by their definition and Assumption \ref{ass1}. By  \cite[Lemma 1]{tibshirani2019conformal}, we hence know  that equation \eqref{eq:pred_direct} is valid by the specific choice of $C$ in equation \eqref{eq:thm_1}. Consequently, we have that 
\begin{align*}
    P\big(\rho^{\phi}(X,\tau_0)\ge \rho^{\phi}(\hat{x},\tau_0)-C\big)\ge 1-\delta.
\end{align*}
If now $\rho^{\phi}(\hat{x},\tau_0)> C$, it holds that $P(\rho^{\phi}(X,\tau_0)>0)\ge 1-\delta$ by which it follows that
\begin{align*}
    P\big((X,\tau_0)\models \phi\big)\ge 1-\delta
\end{align*}
 since $\rho^{\phi}(X,\tau_0)>0$ implies $(X,\tau_0)\models \phi$ \cite{donze2,fainekos2009robustness}.
\end{proof}
\end{theorem}

It is important to note that the direct method, as well as the indirect method presented in the next subsection, do not need to retrain their predictor when the specification $\phi$ changes, as in existing work such as \cite{qin2022statistical,bortolussi2019neural}. This is since we use trajectory predictors to obtain state predictions $\hat{x}_{\tau|t}$ that are specification independent.

\begin{remark}
Note that Theorem \ref{thm:1} assumes a fixed failure probability $\delta$. If one wants to find the tightest bound with the smallest failure probability $\delta$  so that $P\big((X,\tau_0)\models \phi\big)\ge 1-\delta$ holds, we can (approximately) find the smallest such  $\delta$ by a simple grid search over $\delta\in(0,1)$ and repeatedly invoke Theorem \ref{thm:1}.
\end{remark}

\begin{remark}\label{rem_cov_marg}
We emphasize that the prediction regions in equation \eqref{eq:pred_direct}, and hence the result that $P\big((X,\tau_0)>0\models \phi\big)\ge 1-\delta$ if $\rho^{\phi}(\hat{x},\tau_0)> C$,  guarantee marginal coverage. This means that the probability measure $P$ is defined over the randomness of the test trajectory $X$ and the randomness of the calibration trajectories in $D_\text{cal}$.  We thereby obtain probabilistic guarantees for the verification procedure, but we do not obtain guarantees  conditional on $D_\text{cal}$.
\end{remark}

\subsection{Indirect STL Predictive Runtime Verification}

We now present the indirect method where we first obtain prediction regions for the state predictions $\hat{x}_{t+1|t},\hdots,\hat{x}_{t+H|t}$, and then use these prediction regions to solve Problem~\ref{prob1}. We later discuss advantages and disadvantages between the direct and the indirect method (see Remark \ref{remm3}), and compare them in simulations (see Section \ref{sec:simulations}).

For a failure probability of ${\delta}\in(0,1)$, our first goal is  to construct prediction regions defined by  constants $C_\tau$ so that
\begin{align}\label{eq:pred_state}
    P\big(\|X_\tau-\hat{x}_{\tau|t}\|\le C_\tau, \; \forall \tau\in\{t+1,\hdots,t+H\}\big)\ge 1-{\delta},
\end{align}
i.e., $C_\tau$ should be such that the state $X_\tau$ is $C_\tau$-close to our predictions $\hat{x}_{\tau|t}$ for all relevant times $\tau\in\{t+1,\hdots,t+H\}$ with a probability of at least $1-{\delta}$. Let us thus  consider the following nonconformity score that we  compute for each calibration trajectory $x^{(i)}\in D_\text{cal}$ as 
\begin{align*}
    R_\tau^{(i)}:=\|x_\tau^{(i)}-\hat{x}_{\tau|t}^{(i)}\|
\end{align*}
where we recall that $\hat{x}_{\tau|t}^{(i)}$ is the prediction obtained from the observed calibration trajectory $x_{\text{obs}}^{(i)}$. A large nonconformity score indicates that the state predictions $\hat{x}_{\tau|t}^{(i)}$ of $x_\tau^{(i)}$ are not accurate, while a small score indicates accurate predictions. Assume again that the values of $R_\tau^{(i)}$ are sorted in non-decreasing order and define $R_\tau^{(|D_\text{cal}|+1)}:=\infty$ as the $(|D_\text{cal}|+1)$th value. To obtain the values of $C_\tau$ that make equation \eqref{eq:pred_state} valid, we use the results from  \cite{stankeviciute2021conformal,lindemann2022safe}.

\begin{lemma}[\cite{stankeviciute2021conformal,lindemann2022safe}]\label{lem:1}
Given a distribution $(X_0,X_1,\hdots)\sim\mathcal{D}$, the current time $t$, the observations $x_{\text{obs}}:=(x_0,\hdots,x_t)$, the dataset $D_\text{cal}$, and a failure probability $\delta\in(0,1)$. Then the prediction regions in equation  \eqref{eq:pred_state} are valid with $C_\tau$ defined as
\begin{align}\label{eq:lem_1}
    C_\tau:=R_\tau^{(p)}\;\;\text{where}\;\;  p:=\big\lceil (|D_\text{cal}|+1)(1-\bar{\delta})\big\rceil \text{ and } \bar{\delta}:=\delta/H.
\end{align}
\end{lemma}

Note  the scaling of $\delta$ by the inverse of $H$, as expressed in $\bar{\delta}$. Consequently, the constants $C_\tau$ increase with increasing prediction horizon $H$, i.e., with larger formula length $L^\phi$, as larger $H$ result in smaller $\bar{\delta}$ and consequently in larger $p$ according to \eqref{eq:lem_1}. 
 
We can now use the prediction regions of the predictions $\hat{x}_{\tau|t}$ from equation \eqref{eq:pred_state} to obtain prediction regions for $\rho^{\phi}(X,\tau_0)$ to solve Problem~\ref{prob1}. The main idea is to calculate the worst case of the robust semantics $\rho^\phi$ over these prediction regions. To be able to do so,  we assume that the formula $\phi$ is in positive normal form, i.e., that the formula $\phi$ contains no negations. This is without loss of generality as every STL formula $\phi$ can be re-written in positive normal form, see e.g., \cite{sadraddini2015robust}. Let us next define a worst case version $\bar{\rho}^{\phi}$ of the robust semantics $\rho^\phi$  that incorporates the prediction regions from equation \eqref{eq:pred_state}.  For predicates $\mu$, we define these semantics as
 \begin{align*}
     \bar{\rho}^{\mu}(\hat{x},\tau)& := 
     \begin{cases}
     h(x_\tau) &\text{ if } \tau\le t\\
     \inf_{\zeta\in \mathcal{B}_{\tau}} h(\zeta) &\text{ otherwise }
     \end{cases}
 \end{align*}
  where we recall  the definition of the predicted trajectory $\hat{x}$ in equation \eqref{eq:pred_traj} and where $\mathcal{B}_{\tau}:=\{\zeta\in\mathbb{R}^n|\|\zeta-\hat{x}_{\tau|t}\|\le C_\tau\}$ is a ball of size $C_\tau$ centered around the prediction $\hat{x}_{\tau|t}$, i.e., $\mathcal{B}_{\tau}$ defines the set of states within the prediction region at time $\tau$. The intuition behind this definition is that we know the value of the robust semantics ${\rho}^{\mu}(X,\tau)=\bar{\rho}^{\mu}(\hat{x},\tau)$ if $\tau\le t$ since $x_\tau$ is known. For times $\tau>t$, we know  that $X_\tau \in \mathcal{B}_{\tau}$ holds with a probability of at least $1-\delta$ by Lemma \ref{lem:1} so that we compute $\bar{\rho}^{\mu}(\hat{x},\tau):=\inf_{\zeta\in \mathcal{B}_{\tau}} h(\zeta)$ to obtain a lower bound for $\rho^\mu(X,\tau)$ with a probability of at least $1-\delta$. 
  \begin{remark}
  For convex predicate functions $h$, computing $\inf_{\zeta\in \mathcal{B}_{\tau}} h(\zeta)$ is a convex optimization problem that can efficiently  be solved. However, note that the optimization problem $\inf_{\zeta\in \mathcal{B}_{\tau}} h(\zeta)$ may need to be solved for different times $\tau$ and for multiple predicate functions $h$. For non-convex functions $h$, we can obtain lower bounds of $\inf_{\zeta\in \mathcal{B}_{\tau}} h(\zeta)$ that we can use instead. Particularly, let $L_h$ be the Lipschitz constant of $h$, i.e., let $|h(\zeta)-h(\hat{x}_{\tau|t})|\le L\|\zeta-\hat{x}_{\tau|t}\|$. Then, we know that
  \begin{align*}
      \inf_{\zeta\in \mathcal{B}_{\tau}} h(\zeta)\ge h(\hat{x}_{\tau|t})-L_hC_{\tau}.
  \end{align*}
  For instance, the constraint $h(\zeta):=\|\zeta_1-\zeta_2\|-0.5$, which can encode collision avoidance constraints, has Lipschitz constant one.
  \end{remark}

  The worst case robust semantics $\bar{\rho}^\phi$ for the remaining operators (\text{True}, conjunctions, until, and since) are defined in the standard way, i.e., the same way as for the robust semantics $\rho^\phi$, and are summarized in Appendix \ref{app:STL} for convenience. We can now use the  worst case robust semantics to solve Problem \ref{prob1}.

\begin{theorem}\label{thm:2}
Let the conditions of Lemma \ref{lem:1} hold. Given a bounded STL formula $\phi$ in positive normal form that is enabled at $\tau_0$. Then it holds that $P\big((X,\tau_0)\models \phi\big)\ge 1-\delta$ if $\bar{\rho}^{\phi}(\hat{x},\tau_0)> 0$.
\begin{proof}
Note first that $X_\tau\in\mathcal{B}_\tau$ for all times $\tau\in \{t+1,\hdots,t+H\}$ with a probability of at least $1-\delta$ by Lemma \ref{lem:1}. For all predicates $\mu$ in the STL formula $\phi$ and for all times $\tau\in\{0,\hdots,t+H\}$, it hence holds that ${\rho}^\mu(X,\tau)\ge \bar{\rho}^\mu(\hat{x},\tau)$ with a probability of at least $1-\delta$ by the definition of $\bar{\rho}^\mu$. Since the formula $\phi$ does not contain negations\footnote{Negations would in fact flip the inequality in an unfavorable direction, e.g., for $\neg\mu$ it would hold that ${\rho}^{\neg\mu}(X,\tau)\le \bar{\rho}^{\neg\mu}(\hat{x},\tau)$ with a probability of at least $1-\delta$.}, it is straightforward to show (inductively on the structure of $\phi$) that ${\rho}^\phi(X,\tau)\ge \bar{\rho}^\phi(\hat{x},\tau)$ with a probability of at least $1-\delta$. Consequently, if $\bar{\rho}^\phi(\hat{x},\tau)>0$, it holds that $P\big((X,\tau_0)\models \phi\big)\ge 1-\delta$  since ${\rho}^{\phi}(X,\tau_0)>0$ implies $(X,\tau_0)\models \phi$ \cite{donze2,fainekos2009robustness}.
\end{proof}
\end{theorem}

Finally, let us point out conceptual differences with respect to the direct STL predictive runtime verification method.

\begin{remark}\label{remm3}
The state prediction regions \eqref{eq:pred_state} obtained in Lemma \ref{lem:1} may lead to conservatism in Theorem \ref{thm:2}, especially for larger prediction horizons $H$ due to the scaling of $\delta$ with the inverse of $H$. In fact, we require larger calibration datasets $D_\text{cal}$ compared to the direct method to achieve $p\le |D_\text{cal}|$ (recall that $C_\tau=\infty$ if $p> |D_\text{cal}|$). On the other hand, the indirect method is more interpretable and allows to identify parts of the formula $\phi$ that may be violated by analyzing the uncertainty of predicates via the worst case robust semantics $\bar{\rho}^\mu(\hat{x},\tau)$. This information may be helpful and can be used subsequently in a decision making context for plan reconfiguration.
\end{remark}

\section{Case Studies}
\label{sec:simulations}

\begin{figure*}
\centering
\includegraphics[scale=0.15]{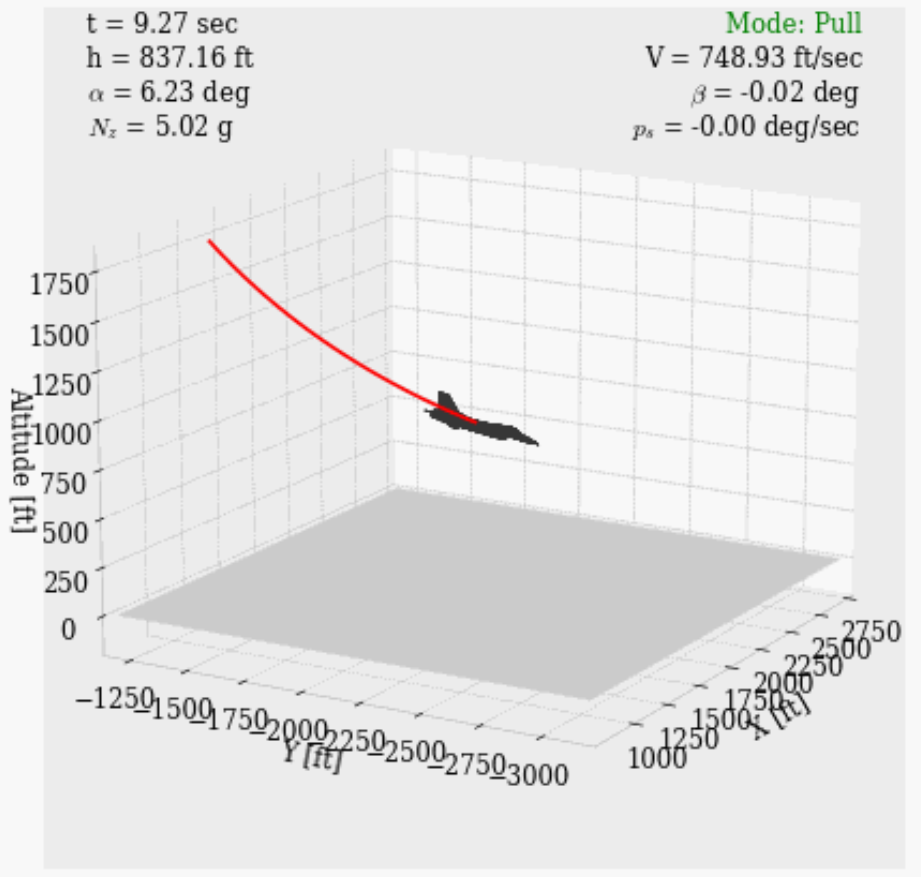}\hspace{1cm}
\includegraphics[scale=0.12]{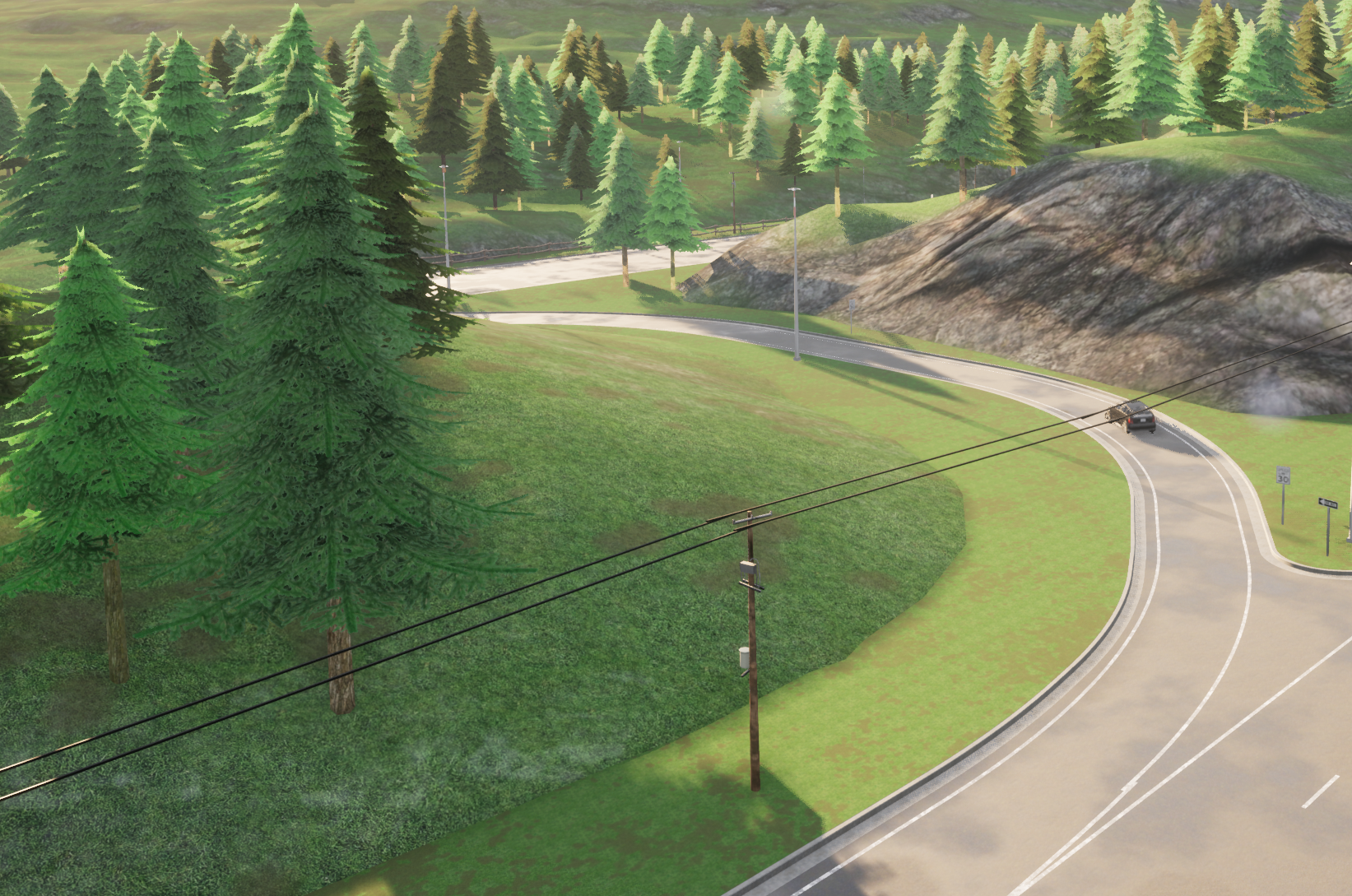}
\caption{Left: F-16 Fighting Falcon within the high fidelity aircraft simulator from \cite{heidlauf2018verification}. Right: Self-driving car within the autonomous driving simulator CARLA \cite{dosovitskiy2017carla}.}
\label{fig:2}
\end{figure*}

To corroborate and illustrate our proposed method, we present two case studies in which we verify an aircraft and a self-driving car. We remark upfront that, in both case studies, we fix the calibration dataset $D_\text{val}$ a-priori and then  evaluate our proposed runtime verification method on several test trajectories. As eluded to in Remark \ref{rem_cov_marg}, one would technically have to resample a calibration dataset for each test trajectory. This is impractical and, in fact, shown to not be needed when the size of the  calibration dataset is large enough, see \cite[Section 3.3]{angelopoulos2021gentle} for a detailed discussion on this topic.

\subsection{F-16 Aircraft Simulator}

In our first case study, we consider the F-16 Fighting Falcon which is a highly-maneuverable aircraft. The F-16 has been used as a verification benchmark, and the authors in \cite{heidlauf2018verification} provide a high-fidelity simulator for various maneuvers such as ground collision avoidance, see Fig. \ref{fig:2} (left). The F-16 aircraft is modeled with $6$ degrees of freedom nonlinear equations of motion, and the aircraft control system consists of an outer and an inner control-loop. The outer loop encodes the logic of the maneuver in a finite state automaton and provides reference trajectories to the inner loop. In the inner loop, the aircraft (modeled by $13$ continuous states) is controlled by low-level integral tracking controllers (adding $3$ additional continuous states), we refer the reader to \cite{heidlauf2018verification} for details. In the simulator, we introduce randomness by uniformly sampling the initial conditions of the air speed, angle of attack, angle of sideslip, roll, pitch, yaw, roll rate, pitch rate, yaw rate, and altitude from a compact set.

\begin{figure*}
\centering
\includegraphics[scale=0.35]{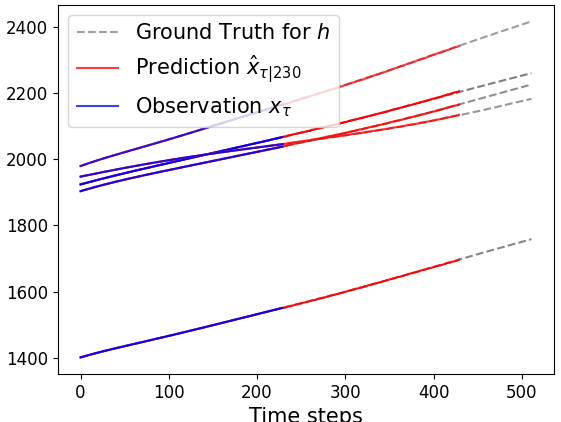}\hspace{1cm}
\includegraphics[scale=0.35]{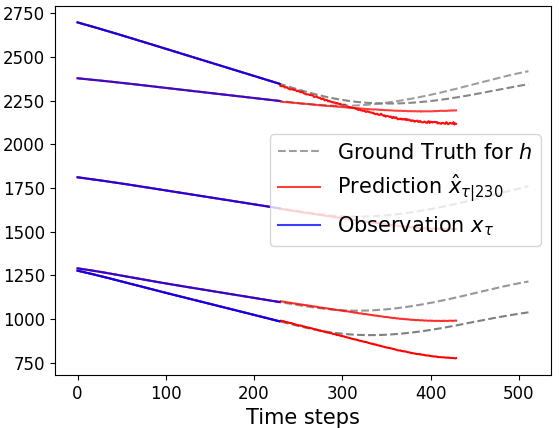}\\
\includegraphics[scale=0.35]{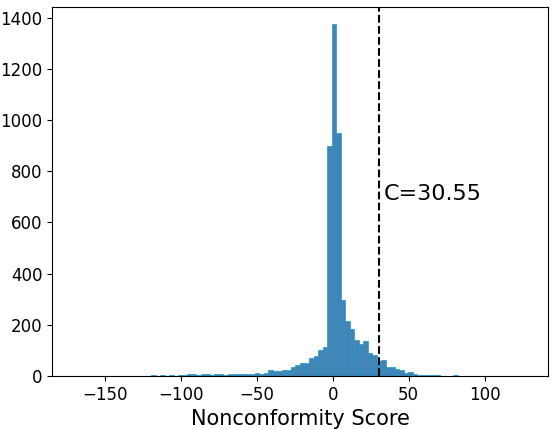}\hspace{1cm}
\includegraphics[scale=0.35]{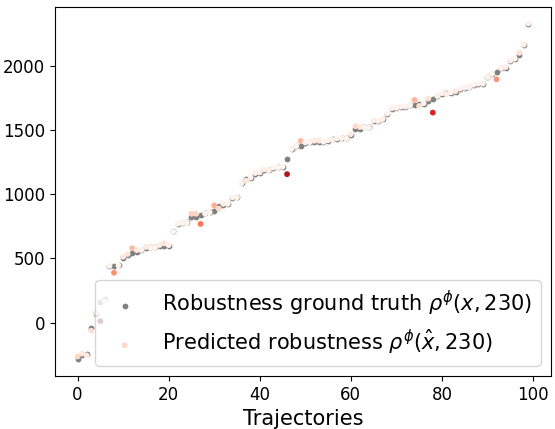}
\caption{LSTM predictions of the altitude $h$ on $D_\text{test}$ (top left, top right) and direct predictive runtime verification method (bottom left, bottom right). Top left: five best (in terms of mean square error) predictions on $D_\text{test}$, top right: five worst predictions on $D_\text{test}$, bottom left: histogram of the nonconformal score $R^{(i)}$ on $D_\text{cal}$ for  direct method, bottom right: predicted robustness $\rho^{\phi}(\hat{x}^{(i)},\tau_0)$ and ground truth robustness $\rho^{\phi}(x^{(i)},\tau_0)$ on $D_\text{test}$.}
\label{fig:f16traj}
\end{figure*}

We use the ground collision avoidance maneuver, and are thus primarily interested in the plane's altitude that we denote by $h$. We collected $D_\text{train}:=1520$ training trajectories, $D_\text{cal}:=5680$ calibration trajectories, and $D_\text{test}:=100$ test trajectories. From $D_\text{train}$, we trained an LSTM of depth two and width $50$ to predict future states of $h$.\footnote{We only used the observed sequence of altitudes  $(h_0,\hdots,h_t)$ as the input of the LSTM. Additionally using other states is possible and can improve prediction performance.} We show the LSTM performance in predicting $h$ in Figure \ref{fig:f16traj}. Particularly, we show plots of the best five
and the worst five LSTM predictions, in terms of the mean square
error, on the test trajectories $D_\text{test}$ in Figure \ref{fig:f16traj} (top left and top right).

We are interested in a safety specification expressed as $\phi:=G_{[0,T]} (h\ge 750)$ that is enabled at time $\tau_0:=t$, i.e., a specification that is imposed online during runtime. Hereby, we intend to monitor if the airplane dips below $750$ meters within the next $T:=200$ time steps (the sampling frequency is $100$ Hz). Additionally, we set $\delta:= 0.05$ and fix the current time to $t:= 230$. 

Let us first use the direct predictive runtime verification algorithm and obtain prediction regions of $\rho^{\phi}(\hat{x},\tau_0)-\rho^{\phi}(X,\tau_0)$ by calculating $C$ according to Theorem~\ref{thm:1}. We show the histograms of $R^{(i)}$ over the calibration data $D_\text{cal}$ in Figure \ref{fig:f16traj} (bottom left). The prediction regions $C$ (i.e., the $R^{(p)}$th nonconformity score) are highlighted as vertical lines. In a next step, we empirically evaluate the results of Theorem \ref{thm:1} by using the test trajectories $D_\text{test}$. In Figure \ref{fig:f16traj} (bottom right), we plot the predicted robustness $\rho^{\phi}(\hat{x}^{(i)},\tau_0)$ and the ground truth robustness $\rho^{\phi}(x^{(i)},\tau_0)$. We found that for $100$ of the $100=|D_\text{test}|$ trajectories it holds that $\rho^{\phi}(\hat{x}^{(i)},\tau_0)> C$ implies
  $(x^{(i)},\tau_0)\models \phi$, confirming Theorem \ref{thm:1}. We also validated equation \eqref{eq:pred_direct} and found that $96/100$ trajectories  satisfy $\rho^{\phi}(\hat{x}^{(i)},\tau_0)-\rho^{\phi}(x^{(i)},\tau_0)\le C$.

\begin{figure*}
\centering
\includegraphics[scale=0.35]{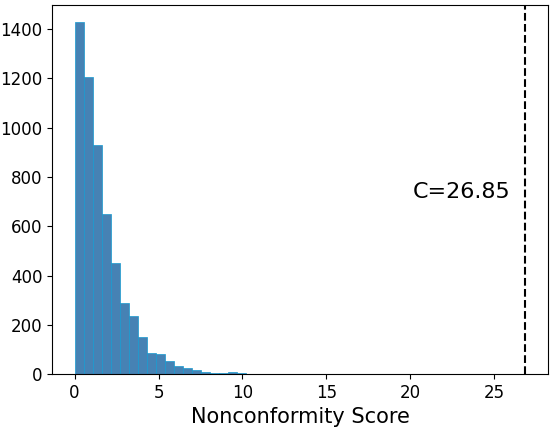}\hspace{1cm}
\includegraphics[scale=0.35]{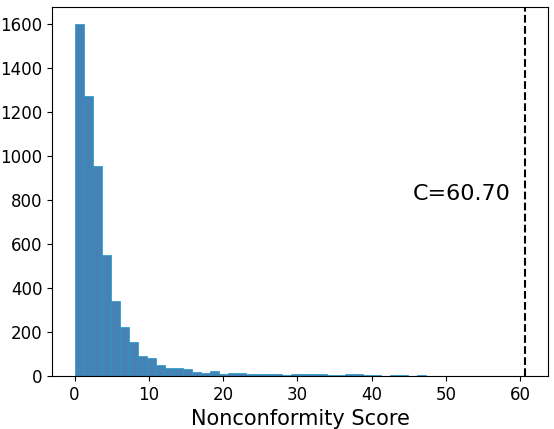}\\
\includegraphics[scale=0.35]{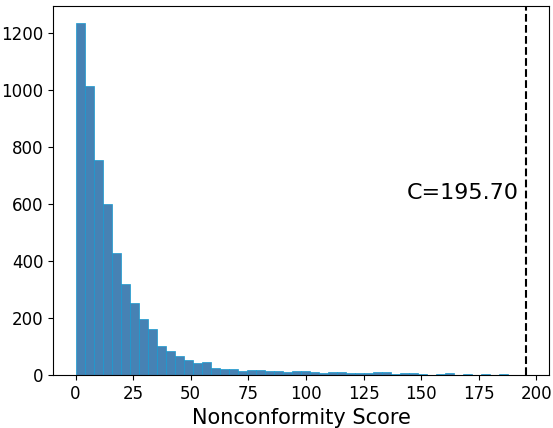}\hspace{1cm}
\includegraphics[scale=0.35]{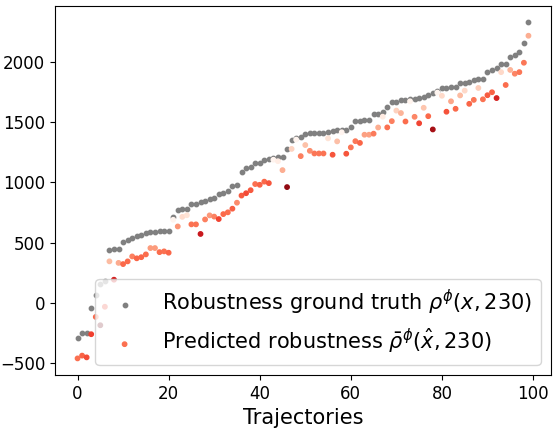}
\caption{Indirect predictive runtime verification method. Top left, top right, and bottom left: histograms of the nonconformal scores $R^{(i)}$ of $\tau$ step ahead prediction on $D_\text{cal}$ for $\tau\in\{50,100,200\}$ and the indirect method, bottom right: worst case predicted robustness $\bar{\rho}^{\phi}(\hat{x}^{(i)},\tau_0)$ and ground truth robustness $\rho^{\phi}(x^{(i)},\tau_0)$ on $D_\text{test}$.}
\label{fig:f16traj1}
\end{figure*}

Let us now use the indirect predictive runtime verification algorithm. We first obtain prediction regions of $\|X_\tau-\hat{x}_{\tau|t}\|$ by calculating $C_\tau$ according to  Lemma \ref{lem:1}. We show the histograms for three different $\tau$ in Figure \ref{fig:f16traj1} (top left, top right, bottom left). We also indicate the prediction regions $C_\tau$ by vertical lines (note that $\bar{\delta}=\delta/200$ in this case). We can observe that larger prediction times $\tau$ result in larger prediction regions $C_\tau$. This is natural as the trajectory predictor is expected to perform worse for larger $\tau$. In a next step, we empirically evaluate the results of Theorem \ref{thm:2} by calculating the worst case robust semantic $\bar{\rho}^\phi(\hat{x}^{(i)},\tau_0)$ for the test trajectories $D_\text{test}$.  In Figure~\ref{fig:f16traj1} (bottom right), we plot the worst case robustness $\bar{\rho}^{\phi}(\hat{x}^{(i)},\tau_0)$ and the ground truth robustness $\rho^{\phi}(x^{(i)},\tau_0)$. We found that for $100$ of the $100=|D_\text{test}|$ trajectories it holds that $\bar{\rho}^{\phi}(\hat{x}^{(i)},\tau_0)> 0$ implies
  $(x^{(i)},\tau_0)\models \phi$, confirming Theorem \ref{thm:2}.

By a direct comparison of Figures \ref{fig:f16traj} (bottom right) and \ref{fig:f16traj1} (bottom right), we observe that the indirect method is more conservative than the direct method in the obtained robustness estimates. Despite this conservatism, the indirect method allows us to obtain more information in case of failure by inspecting the worst case robust semantics $\bar{\rho}^\phi(\hat{x},\tau_t)$ as previously remarked ins Remark~\ref{remm3}.

\subsection{Autonomous Driving in CARLA}
We consider the case study from \cite{lindemann2022risk} in which two neural network lane keeping controllers are verified within the autonomous driving simulator CARLA \cite{dosovitskiy2017carla} using offline trajectory data. The controllers are supposed to keep the car within the lane during a long $180$ degree left turn, see Figure~\ref{fig:2} (right). The authors in \cite{lindemann2022risk} provide offline probabilistic verification guarantees, and find that not every trajectory satisfies the specification. This motivates our predictive runtime verification approach in which we would like to alert of potential violations of the specification already during runtime.

The first controller is based on  imitation learning (IL) \cite{ross2010efficient} and the second controller is based on a control barrier function (CBF) learned from expert demonstrations \cite{lindemann2021learning}. For the analysis, we consider the cross-track error $c_e$ (deviation of the car from the center of the lane) and the orientation error $\theta_e$ (difference between the orientation of the car and the lane). Within CARLA, the control input of the car is affected by additive Gaussian noise and the initial position of the car is  drawn uniformly from  $(c_e,\theta_e)\in [-1,1]\times[-0.4,0.4]$. We obtained $1000$ trajectories for each controller, and use $|D_\text{train}|:=700$ trajectories to train an LSTM, while we use $|D_\text{cal}|:=200$ trajectories to obtain conformal prediction regions. The remaining $|D_\text{test}|:=100$ trajectories are used for testing.  

\begin{figure*}
\centering
\includegraphics[scale=0.35]{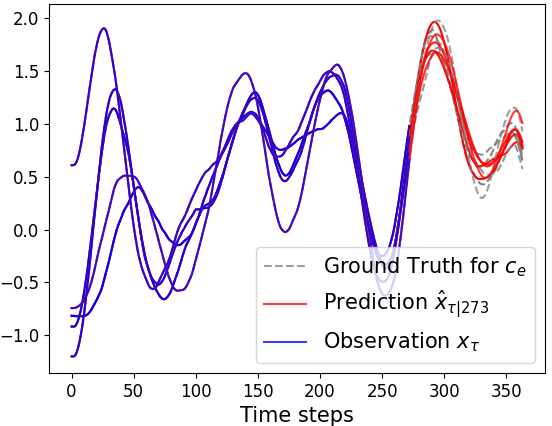}\hspace{1cm}
\includegraphics[scale=0.35]{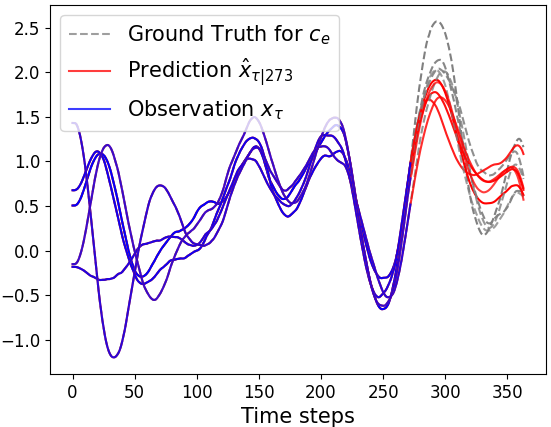}\\
\includegraphics[scale=0.35]{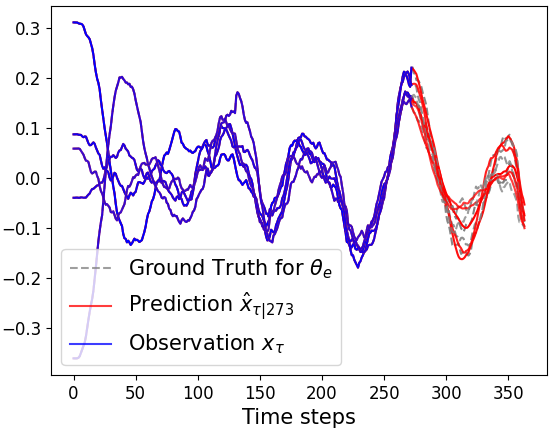}\hspace{1cm}
\includegraphics[scale=0.35]{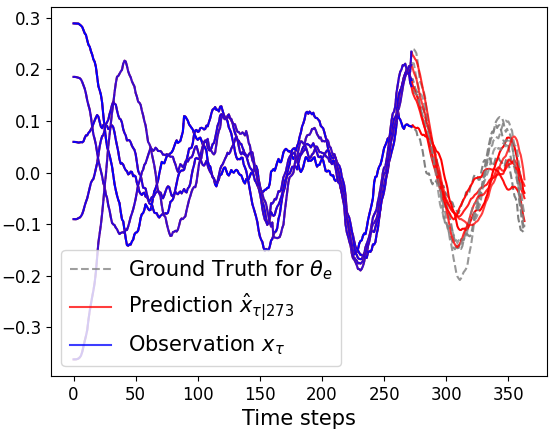}
\caption{LSTM predictions of the imitation learning controller on $D_\text{test}$. Top left: five best (in terms of mean square error) $c_e$ predictions, top right: five worst $c_e$ predictions, bottom left: five best $\theta_e$ predictions, bottom right: five worst $\theta_e$ predictions.}
\label{fig:3}
\end{figure*}

\begin{figure*}
\centering
\includegraphics[scale=0.35]{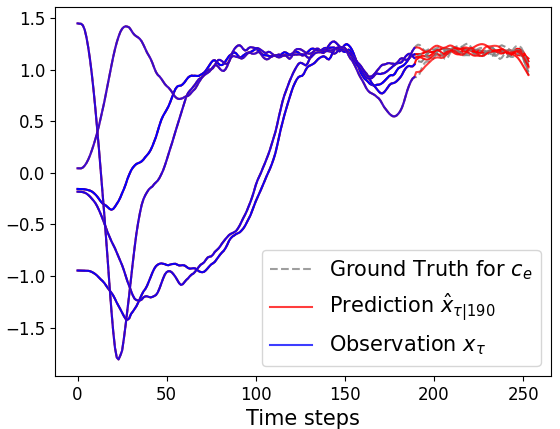}\hspace{1cm}
\includegraphics[scale=0.35]{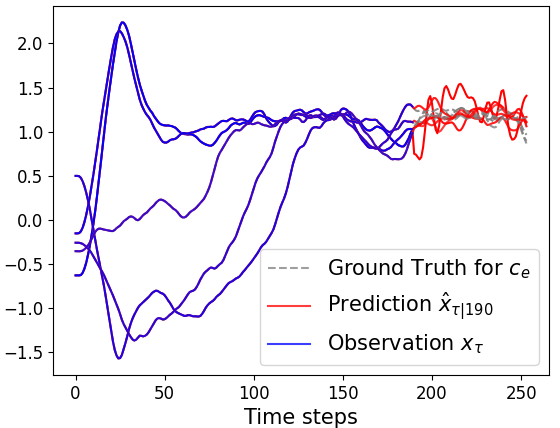}\\
\includegraphics[scale=0.35]{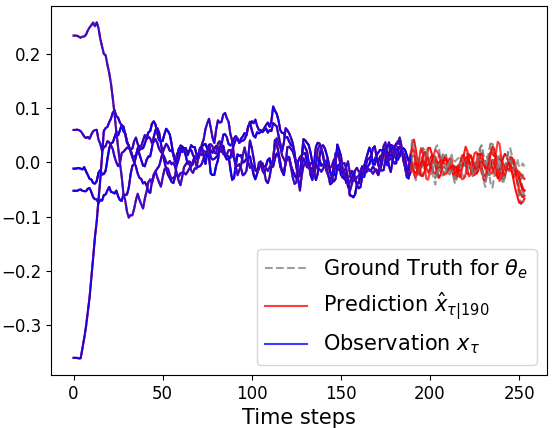}\hspace{1cm}
\includegraphics[scale=0.35]{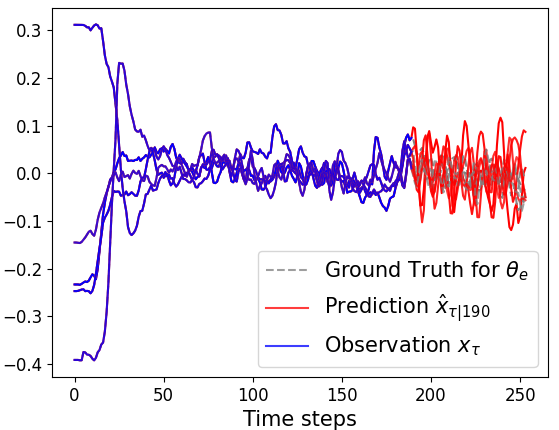}
\caption{LSTM predictions of the control barrier function controller on $D_\text{test}$. Top left: five best (in terms of mean square error) $c_e$ predictions, top right: five worst $c_e$ predictions, bottom left: five best $\theta_e$ predictions, bottom right: five worst $\theta_e$ predictions.}
\label{fig:4}
\end{figure*}

We have trained two LSTMs for each controller from $D_\text{train}$ using the same settings as in the previous section. In Figures \ref{fig:3} and \ref{fig:4}, we show the LSTMs performances in predicting $c_e$ and $\theta_e$ for each controller, respectively. Particularly, the plots show the best five and the worst five LSTM predictions (in terms of the mean square error) on the test trajectories $D_\text{test}$.

 For the verification of the car, we consider the following two STL specifications that are enabled at $\tau_0:=0$:
 \begin{align*}
    \phi_1&:= G_{[10,\infty)]} \big(|c_e|\le 2.25\big),\\
    \phi_2&:= G_{[10,\infty)} \big((|c_e|\ge 1.25) \implies F_{[0,5]}G_{[0,5]}(|c_e|\le 1.25)\big).
\end{align*}
 The first specification is a safety specification that requires the cross-track error to not exceed a threshold of $2.25$ in steady-state (after $10$ seconds of driving). The second specification is a  responsiveness requirement that requires that a cross-track error above $1.25$ is followed immediately within the next $5$ seconds by a phase of $5$ seconds where the cross-track error is below $1.25$. As previously mentioned, we can use the same LSTM for both specifications, and we do not need any retraining when the specification changes which is a major advantage of our method over existing works.

We set $\delta:=0.05$ and fix the current time to $t:=273$ for the IL controller and $t:=190$ for the CBF controller. At these times, the cars controlled by each controller are approximately at the same location in the left turn (this difference is caused by different sampling times).  As we have limited calibration data $D_\text{cal}$ available (CARLA runs in real-time so that data collection is time intensive), we only evaluate the  direct STL predictive runtime verification algorithm for these two specifications.\footnote{The indirect STL predictive runtime verification algorithm would require more calibration data, recall the discussion from Remark \ref{remm3}.} We hence obtain prediction regions of $\rho^{\phi}(\hat{x},\tau_0)-\rho^{\phi}(X,\tau_0)$ for each specification $\phi\in\{\phi_1,\phi_2\}$ by calculating $C$ according to Theorem~\ref{thm:1}. 

\begin{figure*}
\centering
\includegraphics[scale=0.35]{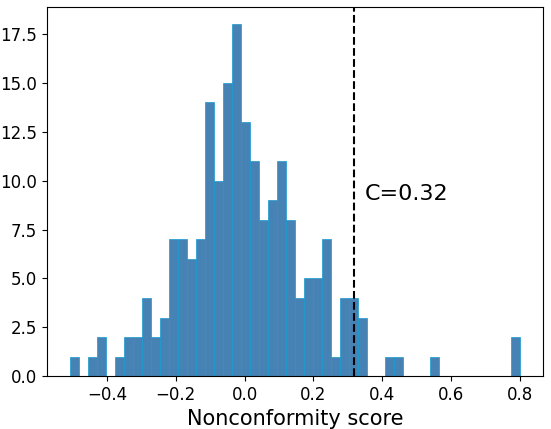}\hspace{1cm}
\includegraphics[scale=0.35]{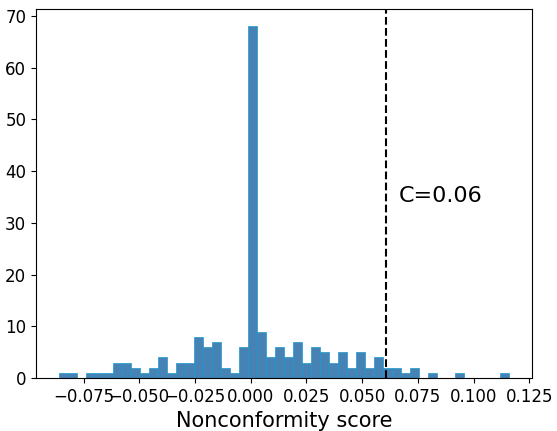}\\
\includegraphics[scale=0.35]{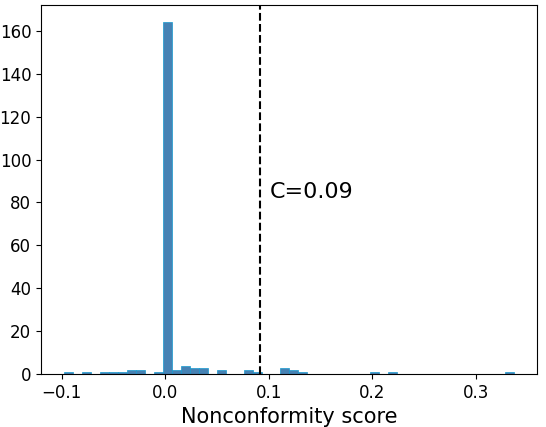}\hspace{1cm}
\includegraphics[scale=0.35]{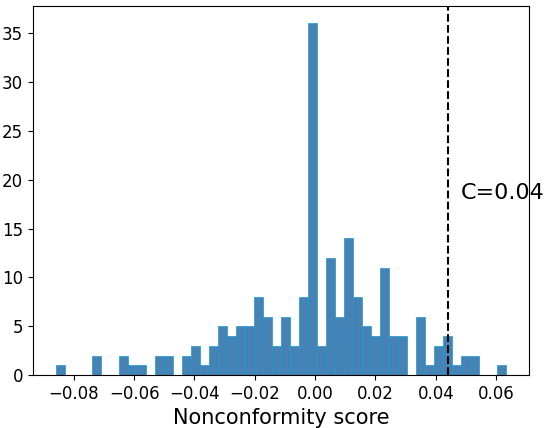}
\caption{Histograms of the nonconformal scores $R^{(i)}$ on $D_\text{cal}$ and prediction region $C$. Top left: IL controller and $\phi_1$, top right: CBF controller and $\phi_1$, bottom left: IL controller and $\phi_2$, bottom right: CBF controller and $\phi_2$.}
\label{fig:5}
\end{figure*}

\begin{figure*}
\centering
\includegraphics[scale=0.35]{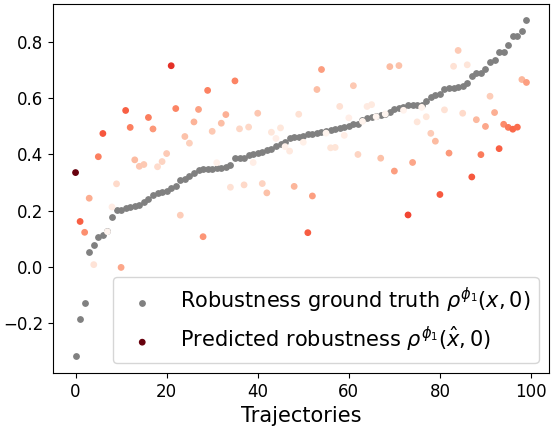}\hspace{1cm}
\includegraphics[scale=0.35]{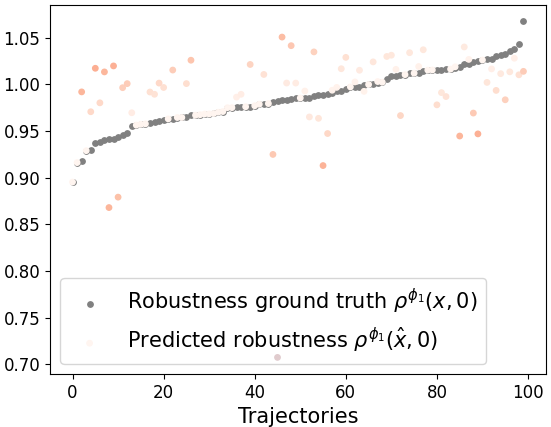}\\
\includegraphics[scale=0.35]{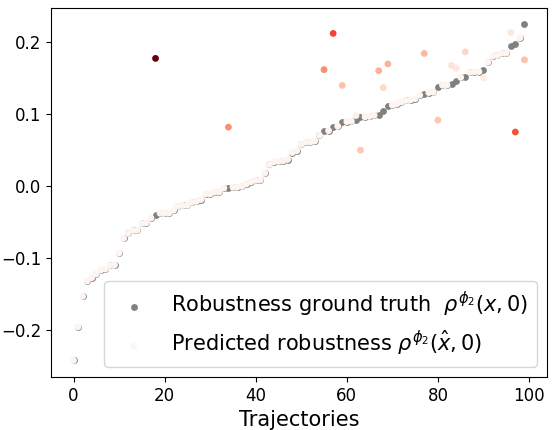}\hspace{1cm}
\includegraphics[scale=0.35]{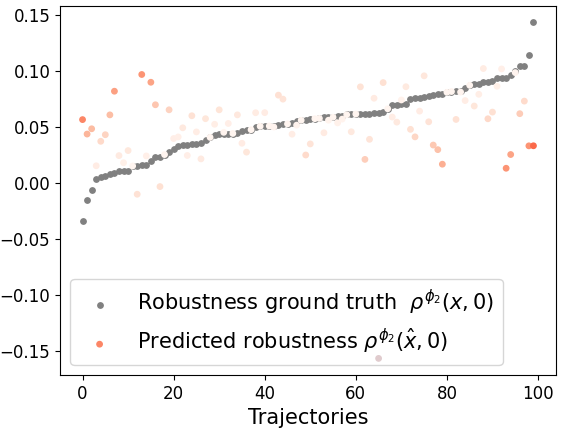}
\caption{Predicted robustness $\rho^{\phi}(\hat{x}^{(i)},\tau_0)$ and ground truth robustness $\rho^{\phi}(x^{(i)},\tau_0)$ on $D_\text{test}$. Top left: IL controller and $\phi_1$, top right: CBF controller and $\phi_1$, bottom left: IL controller and $\phi_2$, bottom right: CBF controller and $\phi_2$.}
\label{fig:6}
\end{figure*}

For the first specification $\phi_1$, we show the histograms of $R^{(i)}$ for both controllers over the calibration data $D_\text{cal}$ in Figure \ref{fig:5} (top left: IL, top right: CBF). The prediction regions $C$ are again highlighted as vertical lines, and we can see that the prediction regions $C$ for the CBF controller are smaller, which may be caused by an LSTM that predicts the system trajectories more accurately (note that the CBF controller causes less variability in $c_e$ which may make it easier to train a good LSTM). In a next step, we empirically evaluate the results of Theorem \ref{thm:1} by using the test trajectories $D_\text{test}$. In Figure \ref{fig:6} (top left: IL, top right: CBF), we plot the predicted robustness $\rho^{\phi_1}(\hat{x},\tau_0)$ and the ground truth robustness $\rho^{\phi_1}(X,\tau_0)$. We found that for $99$ of the $100=|D_\text{test}|$ trajectories under the IL controller and for $100/100$ trajectories under the CBF controller it holds that $\rho^{\phi_1}(\hat{x}^{(i)},\tau_0)> C$ implies
  $(x^{(i)},\tau_0)\models \phi_1$, confirming Theorem \ref{thm:1}. We also validated equation~\eqref{eq:pred_direct} and found that $95/100$ trajectories under the IL controller and $95/100$ trajectories under the CBF controller satisfy $\rho^{\phi_1}(\hat{x}^{(i)},\tau_0)-\rho^{\phi_1}(x^{(i)},\tau_0)\le C$.
  
 For the second specification $\phi_2$, we again show the histograms of $R^{(i)}$ for both controllers over the calibration data $D_\text{cal}$ in Figure \ref{fig:5} (bottom left: IL, bottom right: CBF). We can now observe that the prediction region $C$ for both controllers are relatively small. However, the absolute robustness is also less as in the first specification as can be seen in Figure \ref{fig:6} (bottom left: IL, bottom right: CBF). We again empirically evaluate the results of Theorem \ref{thm:1} by using the test trajectories $D_\text{test}$. In Figure \ref{fig:6} (bottom left: CBF, bottom right: IL), we plot the predicted robustness $\rho^{\phi_2}(\hat{x},\tau_0)$ and the ground truth robustness $\rho^{\phi_2}(X,\tau_0)$. We found that for $99/100$ trajectories under the IL controller and for $98/100$ trajectories under the CBF controller it holds that $\rho^{\phi_2}(\hat{x}^{(i)},\tau_0)> C$ implies
  $(x^{(i)},\tau_0)\models \phi_2$, confirming Theorem \ref{thm:1}. We also validated equation \eqref{eq:pred_direct} and found that $98/100$ trajectories under the IL controller and $92/100$ trajectories under the CBF controller satisfy $\rho^{\phi_2}(\hat{x}^{(i)},\tau_0)-\rho^{\phi_2}(x^{(i)},\tau_0)\le C$.

Finally, we would like to remark that we observed that the added Gaussian random noise on the control signals made the prediction task challenging, but the combination of LSTM and conformal prediction were able to deal with this particular type of randomness. In fact, poorly trained LSTMs lead to larger prediction regions.

\section{Conclusion}
\label{sec:conclusion}

We presented two predictive runtime verification algorithms to compute the probability that the current system trajectory violates a signal temporal logic
specification. Both algorithms use i) trajectory predictors to predict future system states, and ii) conformal prediction to quantify  prediction uncertainty. The use of conformal prediction enables us to obtain valid probabilistic runtime verification guarantees.  To the best of our knowledge, these are the first formal guarantees for a predictive runtime verification algorithm that applies to widely used trajectory predictors such as RNNs and LSTMs, while being computationally simple and making no assumptions on the underlying distribution. An advantage of our approach is that a changing system specification does not require expensive retraining as in existing works. We concluded with  experiments of an F-16 aircraft and a self-driving car equipped with LSTMs.

\section*{Acknowledgements}
Lars Lindemann and George J. Pappas were generously supported by NSF award CPS-2038873. Xin Qin and Jyotirmoy V. Deshmukh gratefully acknowledge the support by the National Science Foundation through the following grants: SHF-1910088, CAREER award (SHF-2048094), CNS-1932620, funding by Toyota R\&D through the USC Center for Autonomy and AI, funding by Airbus Institute for Engineering Research, and gift funding from Northrop Grumman Aerospace Systems. Finally, the authors would like to thank the anonymous reviewers for their feedback.

\bibliographystyle{IEEEtran}
\bibliography{literature}

\appendix
\section{Semantics of Signal Temporal Logic}
\label{app:STL}
For a signal $x:=(x_0,x_1,\hdots)$, the semantics of an STL formula $\phi$ that is enabled at time $\tau_0$, denoted by $(x,\tau_0)\models \phi$, can be recursively computed based on the structure of $\phi$ using the following rules:
	\begin{align*}
	(x,\tau)\models \text{True} & \hspace{0.5cm} \text{iff} \hspace{0.5cm} \text{True},\\
	(x,\tau)\models \mu & \hspace{0.5cm} \text{iff} \hspace{0.5cm} h(x_\tau)\ge 0,\\
	(x,\tau)\models \neg\phi & \hspace{0.5cm} \text{iff} \hspace{0.5cm} (x,\tau)\not\models \phi,\\
	(x,\tau)\models \phi' \wedge \phi'' & \hspace{0.5cm} \text{iff} \hspace{0.5cm} (x,\tau)\models\phi' \text{ and } (x,\tau)\models\phi'',\\
	(x,\tau)\models \phi' U_I \phi'' & \hspace{0.5cm} \text{iff} \hspace{0.5cm} \exists \tau''\in (\tau\oplus I)\cap \mathbb{N} \text{ s.t. } (x,\tau'')\models\phi''\\
	&\hspace{1.2cm} \text{ and } \forall \tau'\in(\tau,\tau'')\cap \mathbb{N}, (x,\tau)\models\phi',\\
	(x,\tau)\models \phi' \underline{U}_I \phi'' & \hspace{0.5cm} \text{iff} \hspace{0.5cm} \exists \tau''\in (\tau\ominus I)\cap \mathbb{N} \text{ s.t. } (x,\tau'')\models\phi''\\
	& \hspace{1.2cm} \text{ and } \forall \tau'\in(\tau'',\tau)\cap \mathbb{N}, (x,\tau)\models\phi'.
	\end{align*}

The robust semantics $\rho^{\phi}(x,\tau_0)$ provide more information than the semantics $(x,\tau_0)\models \phi$, and indicate how robustly a specification is satisfied or violated. We can again recursively calculate $\rho^{\phi}(x,\tau_0)$ based on the structure of $\phi$ using the following rules:
\begin{align*}
	\rho^\text{True}(x,\tau)& := \infty,\\
	\rho^{\mu}(x,\tau)& := h(x_\tau) \\
	\rho^{\neg\phi}(x,\tau) &:= 	-\rho^{\phi}(x,\tau),\\
	\rho^{\phi' \wedge \phi''}(x,\tau) &:= 	\min(\rho^{\phi'}(x,\tau),\rho^{\phi''}(x,\tau)),\\
	%	\rho^{\phi' \vee \phi''}(x,t) &:= 	\max(\rho^{\phi'}(x,t),\rho^{\phi''}(x,t)),\\
	\rho^{\phi' U_I \phi''}(x,\tau) &:= \underset{\tau''\in (\tau\oplus I)\cap \mathbb{N}}{\text{sup}}  \Big(\min\big(\rho^{\phi''}(x,\tau''),\underset{\tau'\in (\tau,\tau'')\cap \mathbb{N}}{\text{inf}}\rho^{\phi'}(x,\tau') \big)\Big), \\
	\rho^{\phi' \underline{U}_I \phi''}(x,\tau) &:= \underset{\tau''\in (\tau\ominus I)\cap \mathbb{N}}{\text{sup}} \Big( \min\big(\rho^{\phi''}(x,\tau''),  \underset{\tau'\in (\tau'',\tau)\cap \mathbb{N}}{\text{inf}}\rho^{\phi'}(x,\tau') \big)\Big).
	\end{align*}
	
	 The formula length $L^{\neg\phi}$ of a bounded STL formula $\phi$ can be recursively calculated based on the structure of $\phi$ using the following rules:
 \begin{align*}
     L^\text{True}&=L^\mu:=0\\
     L^{\neg\phi}&:=L^\phi\\
     L^{\phi'\wedge\phi''}&:=\max(L^{\phi'},L^{\phi''})\\
     L^{\phi' U_I \phi''}&:=\max \{I\cap \mathbb{N}\}+\max(L^{\phi'},L^{\phi''})\\
     L^{\phi' \underline{U}_I \phi''}&:=\max(L^{\phi'},L^{\phi''}).
 \end{align*}
 
 Lastly, we define the worst case robust semantics $\bar{\rho}^{\phi}(\hat{x},\tau_0)$, which are again recursively defined as follows:
\begin{align*}
	\bar{\rho}^\text{True}(\hat{x},\tau)& := \infty,\\
	\bar{\rho}^{\mu}(\hat{x},\tau)& := 
     \begin{cases}
     h(x_\tau) &\text{ if } \tau\le t\\
     \inf_{\zeta\in \mathcal{B}_{\tau}} h(\zeta) &\text{ otherwise }
     \end{cases} \\
	\bar{\rho}^{\neg\phi}(\hat{x},\tau) &:= 	-\bar{\rho}^{\phi}(\hat{x},\tau),\\
	\bar{\rho}^{\phi' \wedge \phi''}(\hat{x},\tau) &:= 	\min(\bar{\rho}^{\phi'}(\hat{x},\tau),\bar{\rho}^{\phi''}(\hat{x},\tau)),\\
	%	\rho^{\phi' \vee \phi''}(x,t) &:= 	\max(\rho^{\phi'}(x,t),\rho^{\phi''}(x,t)),\\
	\bar{\rho}^{\phi' U_I \phi''}(\hat{x},\tau) &:= \underset{\tau''\in (\tau\oplus I)\cap \mathbb{N}}{\text{sup}}  \Big(\min\big(\bar{\rho}^{\phi''}(\hat{x},\tau''),\underset{\tau'\in (\tau,\tau'')\cap \mathbb{N}}{\text{inf}}\bar{\rho}^{\phi'}(\hat{x},\tau') \big)\Big), \\
	\bar{\rho}^{\phi' \underline{U}_I \phi''}(\hat{x},\tau) &:= \underset{\tau''\in (\tau\ominus I)\cap \mathbb{N}}{\text{sup}} \Big( \min\big(\bar{\rho}^{\phi''}(\hat{x},\tau''),  \underset{\tau'\in (\tau'',\tau)\cap \mathbb{N}}{\text{inf}}\bar{\rho}^{\phi'}(\hat{x},\tau') \big)\Big).
	\end{align*}

\addtolength{\textheight}{-12cm}   % This command serves to balance the column lengths

\end{document}